\documentclass[prl,showpacs,twocolumn]{revtex4}
\usepackage{graphics}
\usepackage{epsfig}
\usepackage{graphicx}

\def\be{\begin{equation}}
\def\bea{\begin{eqnarray}}
\def\bma{\begin{mathletters}}
\def\ee{\end{equation}}
\def\eea{\end{eqnarray}}
\def\ema{\end{mathletters}}

\begin{document}
\author{Vlatko Vedral\footnote{e-mail: v.vedral@leeds.ac.uk}}
\title{Entanglement Production in Non-Equilibrium Thermodynamics}
\affiliation{School of Physics and Astronomy, University of Leeds, Leeds, UK \\and\\National University of Singapore, Singapore}
\date{\today}
\date{\today}

\begin{abstract}
We define and analyse the concept of entanglement production during the evolution of a general quantum mechanical 
dissipative system. While it is important to minimise entropy production in order to achieve thermodynamical efficiency, maximising the rate of change of entanglement is important in quantum information processing. 
Quantitative relations are obtained between entropy and entanglement
productions, under specific assumptions detailed in the text. We apply these to the processes of dephasing and 
decay of correlations between two
initially entangled qubits. Both the Master equation treatment as well as the higher Hilbert space analysis are presented. Our formalism is very general and contains as special cases many reported individual instance of entanglement dynamics, such as, for example, the recently discovered notion of the sudden death of entanglement. 
\end{abstract}

\maketitle

\section{Introduction}

Entanglement has been studied extensively in many body systems in the last five years or
so \cite{Amico}. A formalism has been developed to treat entanglement using the standard methods of 
statistical mechanics. Almost all of the studies have been focused on systems in thermal
equilibrium, both at zero and finite temperatures. Real systems are, however, hardly ever
in thermal equilibrium and at a fixed temperature. The field of non-equilibrium thermodynamics
was developed in the decades between 40s and 70s to deal with such driven systems and laws of their evolution
\cite{Mazur}.  

Theoretical predictions of macroscopic entanglement have also been experimentally corroborated in 
systems such as, for example, grains of salt \cite{Ghosh}. This raises a very interesting possibility, which
I briefly speculated about sometime ago \cite{Vedral-Nature}, that entanglement can and does feature in living systems.
Living systems, however, are composed of large macromolecules of various types and they exist at
high temperatures (about 300 Kelvin plus). Can any entanglement survive under such harsh conditions?

It would seem unlikely that macroscopic thermal entanglement could exist at $300$K, but, it is
very important to remember that living systems are not in equilibrium. They are in fact very much
driven by their environments and continuously change in time. For example, chemical reactions 
in the cell's mitochondria are very far from equilibrium, which is why they can produce and 
supply energy necessary for cell's functioning. It is feasible, therefore, that a system's 
equilibrium state is not entangled, while, for the same system, entanglement gets created 
when this system dynamically approaches equilibrium. Steady states of driven systems could also 
easily be engineered to be entangled. Achieving coherence away from equilibrium is nothing unusual
after all: the phenomenon of population inversion is one such instance and it has been speculated that
the coherent electron transport in cell membranes (as well as photosynthesis) functions in a similar 
way \cite{Frohlich}.

Bio-chemical experiments are now approaching the appropriate levels of sophistication where quantum
features of energy transfer processes can be addressed in greater detail. Remarkable recent results show that
quantum coherence, in the form of quantum energy beats, is present in the energy transfer process in photosynthesis at 
$T=77K$ \cite{Fleming}. The authors compare their findings to a version of Grover's search algorithm \cite{Grover}, where the incoming light excites a number of energy states in the receptors, which then coherently (and rapidly) transfers energy to the most convenient storage state. The authors also do not rule out the possibility of non-local effects, i.e. entanglement, though their experiment was not designed to reveal any such phenomenon.

Studying entanglement in real systems, especially biological ones, impels us therefore to phrase the
whole question of the existence of many-body entanglement within the formalism of non-equilibrium 
thermodynamics. The central quantity in 
thermodynamics is entropy. Whether a certain thermodynamical state can be transformed into another one 
is determined by the difference between their corresponding entropies. But, thermodynamics does not tell us how 
exactly the transformation is to be executed - despite its name, the details of dynamics are not part of thermodynamical description. Dynamics of macroscopic objects immersed in noisy environments belongs to the domain of
non-equilibrium thermodynamics. 

The central quantity in non-equilibrium thermodynamics in not entropy, but the rate of change of entropy,
which is known as entropy production \cite{Mazur,Spohn}. This is perhaps not surprising, since dynamics should be 
described by rates
of change of relevant kinematic quantities and entropy is one such quantity. Here, however, our intention
is to study behavior of entanglement in non-equilibrium and we therefore introduce a quantity we call
entanglement production. The main aim of this work will be to compare the behavior of entropy and 
entanglement productions under very general dissipative evolutions. The hope of the author is that the entropy production in some way constrains the rate of entanglement production. This would be a completely novel contribution to the study of connections between entanglement and thermodynamics \cite{MBP98}. Our approach is important not only in the domain of studying macroscopic quantum physics, but may also be able to shed some new light on the issues related to the quantum measurement (see \cite{VedralPRL}).

\section{Entropy production} 

We begin by analysing the following question. A system, whose Hamiltonian is $H$, is
in contact with a heat bath which interacts with the system driving it into the equilibrium 
thermal state $\rho_T = e^{-\beta H}/Z$ where $Z= tr e^{-\beta H}$ is the partition function. Note that here $H$ will be time independent; however, entropy production can be defined for driven systems as well and most of what we have to say will apply to these more general circumstances. We will point this out whenever appropriate as we proceed in our argument. If the system is at the beginning in some state $\rho$, we ask about the entropy change during the
process in which this state thermalises. 

There are two components to the total entropy change \cite{EL87,Vedralroysoc}, $\Delta S_t$, which 
are usually written as $\Delta S_t = \Delta S_{int} + \Delta S_{ext}$. The first term signifies the internal 
entropy change, which is the change of system's entropy, given by
$\Delta S_{int} = S(\rho_T) - S(\rho)$,
while the second term, the external entropy change, is $\Delta S_{ext}  =  - \sum_k  r_k (tr \rho_T H - \langle r_k|H|r_k\rangle)/T = tr (\rho_T - \rho)\ln \rho_T$,
where $\rho = \sum_k r_k |r_k\rangle\langle r_k|$ is the eigen-decomposition of the density matrix.
This term signifies the entropy increase in the environment. It is derived by calculating
the (expected) heat transfer from the system to the environment divided by the temperature. The total 
entropy change is, therefore, conveniently expressed using the relative entropy as
\begin{equation}  
\Delta S_{t} = S(\rho ||\rho_T) = tr (\rho \ln \rho - \rho \ln \rho_T) 
\end{equation}
This quantity is never negative \cite{Lindblad}, which is an expression of the second law of thermodynamics.
If we look at the continuous change of entropy in time, a naive way of thinking
would suggest to us that the entropy production, $\sigma (t)$, should be given by a derivative
of the above in time, like so
\begin{equation}  
\sigma = - \frac{d}{dt}\Delta S_{t} = - tr (\frac{d}{dt}\rho) (\ln \rho - \ln \rho_T) 
\end{equation}
The negative sign corresponds to the fact that, while the relative entropy generally decreases in time, the entropy production itself should be a positive quantity. Remarkably, upon a more rigorous investigation, this turns out to be the correct expression for entropy production \cite{Spohn}. Since $\rho$ describes the most general evolution of an open system, its (continuous) dynamics is given by the Lindblad master equation which has the following general form ($\hbar=1$) \cite{Lindblad-Master}:
\begin{equation}
  \label{eq:mastereq}
  \dot{\rho}=\frac{1}{i}[H,\rho]-\frac{1}{2}\sum_{k=1}^{n}
  \{\Gamma_k^\dagger \Gamma_k\rho+\rho \Gamma_k^\dagger \Gamma_k\ - 2\Gamma_k \rho\Gamma_k^\dagger\},
\end{equation}
where the dot represents the time derivative, and $\Gamma$'s are the dissipative operators. They do not need to satusfy any special requirements, since the combination through which they enter the Master equation already guarantees the ``physicality" of the whole process. Namely, for a small time interval $\Delta t$, we can describe the time evolution of the density matrix by the following completely positive, trace preserving map (CP-map),
$\rho(t+\Delta t)\approx \sum_{k=0}^{n} W_k \rho(t) W_k^\dagger$, where $W_0=1-i\tilde{H}\Delta t$ and $W_k=\sqrt{\Delta t} \Gamma_k\quad$ ($k\in\{1\dots n\}$) are called the
``no-jump'' and jump operators respectively. $\tilde{H}$ is a non-Hermitian Hamiltonian, given by:
$\tilde{H}=H-\frac{i}{2}\sum_{k=1}^n \Gamma_k^\dagger \Gamma_k$. Note that the operators $W_k$ fulfill the completeness relation $\sum_{k=0}^nW_k^{\dagger} W_k=1$. Because of the non-increase of the relative entropy under general CP-maps \cite{Lindblad}, we can also conclude that $\sigma\geq 0$ holds in general (thereby justifying our insertion of the minus sign in the above definition). 

For completeness, it should be stressed the entropy production rate can be defined even when we do not have stationary states of our evolution. The derivation here followed the assumption that the state of the system thermalises due to its interaction with the environment, but even outside of this framework - and we mentioned driven systems earlier - the concept of entropy production is still very much meaningful. Since in this paper most of our attention will be devoted to dissipative evolutions leading to thermalisation, our above considerations will be sufficient, but it is by all means worth noting that the generality of the concept of entropy production stretches far beyond what we have presented.  

Thermodynamically the most efficient processes are the reversible ones, where $\sigma = 0$. A state of the system must then be transformed by being in touch with (a sequence of) reservoirs whose state is only infinitesimally different to that of the system. Then, the relative entropy between the system and the reservoir is zero up to the second order, $S(\rho+\delta\rho||\rho) \approx 0$. Thermodynamical efficiency, however, may not be the desired goal when we are optimising the computational speed-up, as we do in quantum computation. Then, a far better indicator of efficiency is the rate of consuming or creating entanglement (whether we do the former or the latter depends on the model of quantum computers we use, but this distinction is of little importance here). We now proceed to define it. 

\section{Entanglement Production} 

The first important point to mention is that when a system interacts with its environment, there are many different entanglements that we can consider. For instance, there is the entanglement between the system and the environment that develops as the evolution, in the form of their interaction, proceeds. Secondly, there is the entanglement within the environment itself, which could be very complex. In this paper, however, we are interested in the amount of entanglement within the system only. Much as the entropy production refers to the entropy production by the system, so will the entanglement production be strictly confined to within the system.  

Given this, there is still a myriad different expressions for the amount of entanglement in a given quantum state $\rho$ \cite{Vedral-RMP}. However, since the relative entropy is involved in quantifying the entropy production, we will also use the relative entropy to measure entanglement \cite{Vedraletal,Vedral-Plenio}. Another advantage of this measure is that it is universal, applying as it does to any number of subsystems of any dimensionality \cite{Vedral-RMP}. We define entanglement production, 
$\sigma_E$, to be the temporal derivative of the relative entropy of entanglement. Thus,
\begin{equation}  
\sigma_E = \frac{d}{dt} S(\rho(t)||\rho_{sep} (t))  
\end{equation}
This can be easily rewritten as $\sigma_E = tr \{\dot{\rho}\ln \rho - \exp\{\ln \rho - \ln \rho_{sep}\} \dot{\rho}_{sep}\}$, highlighting the difficulty that, unfortunately, exists in computing the last term, $\dot{\rho}_{sep}$. This is because the closest separable states changes in time with $\rho$ and we need to calculate it at each instant in time. Ignoring this difficulty for the moment (for this already exists when calculating the relative entropy of entanglement itself \cite{Vedral-RMP}), let us first analyse the general relationship between entropy and entanglement productions.

How do we expect $\sigma$ and $\sigma_E$ to be related, given our knowledge of dissipation and entanglement? We would
most likely conjecture that the rate of entropy production is in absolute sense always greater than that of entanglement production. The reason is that we can easily imagine a situation where the steady state is disentangled, and therefore $\sigma_E=0$, but the entropy is still produced, i.e. $\sigma >0$ (see for example \cite{Breuer}). But, does this hold more generally? This simple answer is no, since we can construct a state very weekly coupled to its environment, so that the resulting entropy production is very small, but that its (state's) internal dynamics is so strong that entanglement gets rapidly generated.  Since the rate of internal dynamics is (seemingly) completely independent from the coupling rate to the environment, this difference between the entropy and entanglement productions can be made as large as we please (this is not quite correct since both the driving and dissipation may depend on the same features, but we will not go into details of this here \cite{Plenio-Knight}). What if, on the other hand, the system is only evolving entanglement due to the environmental coupling (i.e. without any other external driving)?

To investigate this, let us now assume a special case where $\rho_{sep} (t) = \rho_T$ at all times (we will encounter a concrete example of such a system later). Then, $d\rho_{sep}/dt = 0$, and so
$\sigma_E (t) + \sigma (t) = 0$. In this case entanglement decreases by exactly as much as the entropy increases at each instant in time. 
Since we know that entropy production is always positive (this is a restatement of the Second Law), the above equation implies that entanglement is being reduced. All the dissipation is therefore used up solely in destroying entanglement. In general, however, this is not necessarily the case. The separable state will in general evolve during the dissipative evolution and the above law no longer holds. What then is the function of the excess of entropy production over entanglement production? 

Before we address this issue, it is interesting to mention Prigogine \cite{Prigogine} who showed that under some restricted conditions, a principle of the minimum entropy production can be derived (for details see \cite{Mazur}). We hear it frequently stated that biological processes conform to this rule, which in our above case would imply the principle of minimum entanglement destruction. It is fascinating to explore further if the efficiency of some natural processes simply derives from the maximal possible preservation of entanglement during the time of these processes. (Might the same be true for general quantum computational processes?).

We argued that it is clearly feasible to have a finite entropy production while at the same time not have any entanglement dynamics. A steady state of a driven system displays this feature as we noted before, but so does the recently experimentally confirmed \cite{ent-death} phenomenon of the ``sudden death of entanglement" \cite{Eberly}. In the latter case, the system dissipates toward a disentangled steady state, but entanglement vanishes before this steady state is reached. Beyond this point, it is clear that $\sigma_E=0$, but entropy is still produced until the steady state is reached. 

The opposite of this cannot happen, namely the fact that entanglement is produced but entropy is not, providing that the system is not driven. If the system is not driven, the entropy production is zero only in thermal equilibrium, but then so is entanglement production. It is therefore natural to conjecture that, if we eliminate the possibility of driven entanglement, then $|\sigma_E| \le \sigma$. We use the absolute value of entanglement production because we do not want
to worry about whether entanglement increases or decreases. We would only like to claim that the change in entanglement
is bounded by the change in entropy. This question resembles the question of Landauer's erasure \cite{Landauer}, namely whether the entropy increase in the environment during a measurement is greater than the information obtained during the same measurement.  In order to address the question of entanglement production in a more quantitative way, we now utilise a different way of presenting dissipative evolutions. 

\section{Higher Hilbert Space View}

Any completely positive map of the above type can be represented in 
a fully unitary way providing we are allowed to include the environment in the treatment of the dynamics. At 
this global level of the system and environment, the dynamics is strictly unitary (since their aggregate is closed). Where does entropy production come from then, when unitarity strictly preserves entropy? The answer 
is that it is generated by neglecting the correlations between the system and the environment \cite{Peres}. We will see this in our examples below, but for now let us imagine that the initial state of the system and environment is uncorrelated, $\rho_S\rho_E$. Suppose that after the unitary interaction the state is $\rho_{S'E'}$ (primes will always pertain to the evolved state). It is true that due to unitarity $S(\rho_S\rho_E) =S(\rho_{S'E'})$, however, if we separate the system and environment then, due to subadditivity, $S(\rho_{S'E'})\le S(\rho_{S'}) + S(\rho_{E'})$. Therefore, if we disregard
the correlations, we have that $S(\rho_S) + S(\rho_E) \le S(\rho_S') + S(\rho_E')$ and so 
$\Delta S_t = \Delta S_S + \Delta S_E = S(\rho_S') - S(\rho_S) + S(\rho_E') + S(\rho_E) \ge 0$. A way of 
explaining the entropy increase from a unitary evolution lies therefore in neglecting correlations (this is
probably the most accepted view of ``deriving" irreversibility from the microscopic reversibility).   

I would like to make a simple point here that is well known, but may be worth discussing briefly. It is this. The view that the deletion of correlations is responsible for the entropy increase is very closely related, if not identical, to the ``coarse graining" method of explaining the entropy increase (that originated in classical physics). The coarse graining argument goes as follows. Any Hamiltonian evolution (be it quantum or classical) preserves entropy and is incapable of explaining its increase. However, if we calculate the entropy with respect to some underlying structure (known as the coarse grained version of phase space in classical physics), then this relative entropy between our state and its coarse grained version will always increase. Here, it is the neglect of the underlying microscopic structure that is responsible for entropy increase and the relative entropy again becomes a prominent measure of the ``amount of neglect". It turns out, furthermore, that the relative entropy between the quantum state $\rho_{AB}$ and its marginals $\rho_A\rho_B$ is just equal to the mutual information of AB, $S(\rho_A) + S(\rho_B) - S(\rho_{AB})$ \cite{Vedral-RMP}. And, we have seen that the ``flow" of relative entropy, unlike that of ordinary entropy, is always unidirectional when we have the most general CP evolution.  It is this change in relative entropy that is used to quantify the entropy production. It is thus immaterial whether we use the mutual information or the relative entropy of coarse graining to measure the entropy increase - they should ultimately be the same.       

Suppose, following this logic, that our system S now contains two subsystems A and B. The initial state of 
the system and envorinment is, as before, $\rho_{AB}\rho_E$. The evolution is given by
$U_{ABE} \rho_{AB}\rho _{E}U_{ABE}^{\dagger} = \rho_{A'B'E'}$. We can now apply the strong 
subadditivity \cite{Ruskai} to the final state to obtain $S_{A'B'E'} + S_{B'}\le S_{A'B'} + S_{B'E'}$.
Since $S_{A'B'E'} = S_{ABE} =S_{AB} +S_E$, we have that $\Delta S_{AB} + \Delta S_E \ge S_{B'} + S_{E'}-S_{B'E'} = I(B',E')$, which means that the entropy increase due to separation is bigger than the mutual 
information between either of the subsystems and the environment (see \cite{Lindblad-JPhys} for a more general discussion of the thermodynamics of measurement information). We have shown previously \cite{Henderson} that, under quite general circumstances, the change in the mutual information between environment and the system is an upper bound to the change of entanglement. (Similar considerations were presented in \cite{Groisman} where the authors show that the amount of work needed to erase all correlations in a state is equal to its mutual information). Following through the above inequality this would imply that $\Delta S_{AB} + \Delta S_E \ge \Delta E_{AB}$; this, inequality, however, still remains a conjecture, although we have presented strong ``circumstantial" evidence in its favor. 

The issues related to entropy and entanglement productions have been very general so far, and hold for any kind of physical system independently of its size, number of degrees of freedom and such. From now on, however, we will specialise to the evolution of two qubits under the influence of dephasing and dissipation. These are two very common mechanisms leading to thermalisation which is why they are of particular importance. We hope, however, that our considerations will, in the future, be extended to many body systems, where - as we already indicated in the introduction - many interesting and fundamental questions are to be found (for a review see \cite{Amico}).  

\section{Examples: dephasing and dissipation}

Suppose that systems $A$ and $B$ are two entangled qubits
initially in an entangled state $|\Psi_{AB} = a\left| 00\right\rangle +b\left|
11\right\rangle$. Here the relative entropy of entanglement is $E_{AB}=-a^2\ln a^2-b^2\ln b^2$ (We assume $a,b$ to be real for simplicity and normalised $a^2+b^2=1$). There are many way of destroying
this entanglement, but the two most common ones are through dephasing or decay. Imagine that one of our qubits
is coupled to an environment leading to a dissipative evolution. At the Master equation level, this can be written as 
\begin{eqnarray}  
\frac{d\rho}{dt} & = & \frac{\omega}{2i} [\sigma_z, \rho] + \frac{\gamma}{2} ([\sigma^{-}\rho, \sigma^{+}] + [\sigma^{-},\rho \sigma^{+}]) \nonumber \\
& + & \frac{\kappa}{2}([\sigma^{+}\rho, \sigma^{-}] + [\sigma^{+},\rho \sigma^{-}]) - \frac{\eta}{2}[\sigma_z,[\sigma_z,\rho]] 
\end{eqnarray}
where $\sigma^{\pm}_z$ are the usual Pauli matrices (instance of the $\Gamma$ matrices in the general formulation) and $\gamma,\kappa,\eta\ge 0$ are decay rates. The first term describes the free qubit evolution under $H=\omega \sigma_z$ (with $\omega$ being its natural frequency), the second and third terms are the decay ones (in both directions of exciting and de-exciting the qubit), while the last term indicates the dephasing process. Here we have assumed that only one of the qubits interacts in this way with its environment, while the other one remains stationary. This restricted assumption is immaterial to our discussion. We could easily have included the evolution of the other qubit without any fundamental modification to the forthcoming conclusions.

It is straightforward to compute the overall state evolution under this Master equation,
$$
\rho (t)=\left(\begin{array}{cccc}
a(t) &0&0&\mu (t)ab \\
0&0&0&0\\
0&0&0&0\\
\mu (t)ab &0&0& 1-a(t) \end{array} \right) \; ,
$$
where $\mu (t) = \exp \{-i\omega t - (1/2)(\gamma+\kappa+\eta)t\}$, $a(t) = \delta (t) a^2 + \delta'(t)$, $\delta (t) = \exp (-(\gamma+\kappa)t)$, and $\delta' (t) = (\gamma (1-\exp (-(\gamma+\kappa)t)))/(\kappa+\gamma)$. The stationary state of this evolution is obtained by letting $t\rightarrow \infty$, while the closest separable state at any instant in time is given by deleting the off-diagonal elements, i.e. by setting $\mu =0$ in the above density matrix. These allow us to compute both the entropy as well as entanglement productions. The expressions are simple but cumbersome to write down; the difference between entropy and entanglement productions is: $\sigma-|\sigma_E| = \dot{a}(t) \ln (1-a(t))a_{\infty}/(a(t)(1-a_{\infty}))$, where $a_{\infty} = \lim_{t\rightarrow\infty} a(t)$. It can be proved that this is a positive quantity (thus supporting our conjecture) that decays at the rate $e^{-(\kappa+\gamma)t}$. Note that the off-diagonal decay does not play any role in this formula, although, of course, it is the reduction of the off-diagonals that contributes to thermalissation (and thereby it features in both entropy and entanglement productions individually).    

We can now investigate the process of dephasing on its own, but to make it clearer we will undertake this from the higher Hilbert state picture. We have that $\kappa=\gamma=0$ and $\eta>0$. Dephasing is obtained through the following interaction between $(AB)$ and $E$, $(\left| 0_{A}0_{B}\right\rangle  +  \left| 1_{A}1_{B}\right\rangle )\left|
0_{E}\right\rangle \rightarrow \left| 0_{A}0_{B}\right\rangle \left|
a_{E}\right\rangle  + \left| 1_{A}1_{B}\right\rangle \left| b_{E}\right\rangle$, 
where $\left\langle a_{E}|b_{E}\right\rangle =1 - \eta\delta t>0$. Suppose that the same interaction happens in 
each small time step $\delta t = t/k$, where $t$ is the total time of evolution. Then, after $k$ such steps we obtain, 
\[
\left| 0_{A}0_{B}\right\rangle \underbrace{\left| a_{E}\right\rangle \left|
a_{F}\right\rangle ...\left| a_{N}\right\rangle}_k+\left|
1_{A}1_{B}\right\rangle \underbrace{\left| b_{E}\right\rangle \left|
b_{F}\right\rangle ...\left| b_{N}\right\rangle}_k
\]
But now $\left\langle a_{E}|b_{E}\right\rangle \left\langle
a_{F}|b_{F}\right\rangle ...\left\langle a_{N}|b_{N}\right\rangle
=(1-\eta t/k)^{k}\rightarrow \exp (-\eta t)$ as $k\rightarrow \infty$. The final state of
the system AB, when the environment is traced out, is the mixture $|00\rangle\langle 00| + |11\rangle\langle 11|$, 
which is exactly the same result as that obtained from the Master equation treatment above. This is not
an accident; this way of treating the dephasing is equivalent to the two assumptions
used in deriving the Master equation, the so-called Born and Markov
approximations. The former ensures that during each step, $\delta t$, the interaction with the environment is 
weak (i.e. $\eta\delta t$ is ``small"), while the latter effectively corresponds to a memoryless 
environment (each new interaction is with a different environmental subsystem). In this
case the state of $AB$ becomes mixed as $t\rightarrow \infty$ and contains no entanglement. 
Quantum correlations have thus been destroyed, but the classical ones still remain. Note that the closest separable state throughout the evolution is always the same and it is equal to the final state reached as $k\rightarrow \infty$. 

An interesting aspect of this treatment is that the mutual information between the system and environment, whose total state is pure, is twice the value of the entropy of the state, and, twice the value of the reduction of entatanglement. This seems to represent a contradiction in the light of the previous claim, namely that when the thermal and separable state coincide, then the entanglement reduction should be equal to entropy production. However, the environment in this example was pure to start with. In order to recover the equality of entropy and entanglement rates, we need to start with an already mixed environment (many identical subsystems, all completely dephased). Then, through the interaction with the system, dephasing will be achieved whose total entropy increase will exactly match the entanglement decrease. This shows that a more appropriate way of looking at dephasing is through the process of thermalisation as presented in \cite{Scarani}. 

\section{Concluding discussion} 

Here we have introduced the notion of entanglement production, in direct analogy with the quantity called entropy production. The latter has been used extensively in the studies of non-equilibrium thermodynamics leading to fundamental and profound insights in chemistry and biology \cite{Coveney}. We have shown that the two quantities are closely related, and have presented evidence for the fact that entropy production presents a bound on the rate of change of entanglement during general dissipative processes. 

There are many open question of general nature implied by our work. For instance, is there an entanglement ``balance" equation of the type 
$dE/dt = \sigma_E + \nabla j_E$,
where $\nabla j_E$ would be the entanglement flux of entanglement through the boundary between the system and its environment? It is clear that this can be answered in the affirmative in the case when $\sigma_E=\sigma$, since entropy production satisfies a balance equation, but does it hold more generally? 

While entropy is important in describing the directionality as well as the speed of thermal processes, entanglement and the rate of its change are important when we discuss the capacity of the system for quantum information processing. One wonders, for instance, if the efficiency of one way quantum computers relys on a delicate balance and trade-off between entropy production generated by measusments and the entanglement reduction during computation \cite{Anders}. In a broader context, since we are discovering more and more rapidly that natural processes exploit quantum effects to enhance their efficiency, studying general principles behind dissipative entanglement production appears to be a very important and worthwhile venture.

\noindent {\bf Acknowledgments.} The author would like to thank the Royal Society and the Wolfson Trust as well as
the Engineering and Physical Sciences Research Council for their financial support.

\end{document}